\documentclass[11pt,a4paper]{article}
\usepackage{epsfig}
\usepackage{amsmath}

\newcommand{\beq}{\begin{equation}}
\newcommand{\eeq}{\end{equation}}
\newcommand{\bea}{\begin{eqnarray}}
\newcommand{\eea}{\end{eqnarray}}

\begin{document}
\title{Broken promises and quantum algorithms}
\author{Adam Brazier and Martin B. Plenio \\ QOLS, Blackett Laboratory, \\Imperial College, \\Prince Consort Road, London, SW7 2BW, UK}
\maketitle
\begin{abstract}In the black-box model, problems constrained by a `promise' are the
only ones that admit a quantum exponential speedup over the best classical algorithm in terms of
query complexity. The most prominent example of this is the Deutsch-Jozsa algorithm. More recently,
Wim van Dam put forward an algorithm for unstructured problems (i.e., those without a promise). We
consider the Deutsch-Jozsa algorithm with a less restrictive (or `broken') promise and study the
transition to an unstructured problem. We compare this to the success of van Dam's algorithm. These
are both compared with a standard classical sampling algorithm. The Deutsch-Jozsa algorithm remains
good as the problem initially becomes less structured, but the van Dam algorithm can be adapted so
as to become superior to the Deutsch-Jozsa algorithm as the promise is weakened.
\end{abstract}

\section{Introduction}

It is known that for a quantum algorithm to achieve a black-box exponential speedup over a
classical one, the problem in question must be constrained by a `promise' (Buhrman et al.,
\cite{bbcmd:ql}). In the Black Box model of computation, where the oracle contains a set of $N$
Boolean variables $F = (f_{0}, f_{1}, f_{2} \ldots f_{N-1})$, about the properties of which
questions can be asked (eg, `do all the $f_{i}$ have the value 1', or `do the $f_{i}$ contain at
least one 1') through queries of the oracle, a problem with a promise is one with a restriction on
the $F$s that are allowed.

The Deutsch-Jozsa problem is an example of a problem with a promise, where the question is one of
deciding whether or not a given function is `balanced' or `constant' (each of these criteria
describes a set of possible $F$ from the set of all possible $N$-bit strings and so comprises a
promise as described above). We consider a relaxation of the promise on this problem (i.e., we
allow extra possible $F$) and consider how this affects the efficacy of the Deutsch-Jozsa algorithm
as the weakening of the promise is increased and the problem becomes less structured.

Following an introduction to the standard Deutsch-Jozsa problem (in section \ref{sec:DJpromise}),
we modify/weaken the promise on the Deutsch-Jozsa problem in section \ref{sec:brokprom} and then
consider the performance of the Deutsch-Jozsa algorithm on this modified problem, particularly in
the limit of large numbers of input qubits and queries, providing asymptotic results. A classical
algorithm based on sampling is devised to address the new problem and its success probability is
considered in the same limits. In section \ref{sec:wvd} we introduce van Dam's Quantum Oracle
Interrogation algorithm \cite{wvd:qo} (which is designed for \emph{unstructured} problems),
consider how well it can be adapted to solve the modified problem and compare its performance with
that of the Deutsch-Jozsa algorithm as the problem becomes less structured; this allows us to
determine in which regimes one might prefer to use which algorithm.

\section{The Deutsch-Jozsa Problem and Algorithm}\label{sec:DJpromise}

The Deutsch-Josza problem is to recognise whether a Boolean function $f$, $\{ 0,1
\}^{n}\stackrel{f}{\mapsto} \{ 0,1 \}$ is `balanced' or `constant' (that the function has one of
these two properties is the promise on the problem). A constant function is one for which $f(x)$
evaluates the same, for all $x \in \{ 0,1 \}^{n}$ (i.e., either the $f(x)$ are all $1$s or all
$0$s). A balanced function is one in which $f(x)$ is equal to $0$ for exactly half of the $x$ and
$1$ for the other half. This comprises a restriction on the possible contents of the oracle,
$F=f_{0}f_{1}\ldots f_{N-1}$ where $N=2^{n}$ and $f_{x}=f(x)$. The Deutsch-Jozsa algorithm
\cite{d:dj}-\cite{cemm:qar} solves this problem with one query and with no probability of error.

The input state for the Deutsch-Josza algorithm is an equal superposition of all the $|x\rangle$
such that $x \in \{ 0,1 \}^{n}$ (constructed by acting on $|00 \ldots 0\rangle$ with $H^{\otimes
n}$), with an ancilla qubit in the state $\frac{1}{\sqrt{2}}(|0\rangle - |1\rangle)$. This is then
acted on with the operator $U_{f}: U_{f}|x \rangle |y \mapsto |x \rangle |y \oplus f(x) \rangle$
which, given that $|y \rangle$ is the ancilla bit already described and `$\oplus$' represents
addition modulo $2$, has the effect of introducing a phase $(-1)^{f(x)}$ to the state $|x \rangle$.
Finally an $n$-bit Hadamard is applies to the first $n$ qubits (the ancilla is discarded) and then
the resulting state is measured along $|z=0 \rangle$. The process, omitting the ancilla quibit, is
summarised below:

\begin{eqnarray} |00 \ldots 0 \rangle &\stackrel{H^{\otimes n}}{\mapsto}& \frac{1}{\sqrt{2^{n}}} \sum_{x=0}
^{x=2^{n}-1} |x \rangle \nonumber \\ &\stackrel{U_{f}}{\mapsto}& \frac{1}{\sqrt{2^{n}}} \sum_{x=0}
^{x=2^{n}-1} (-1)^{f(x)}|x \rangle \nonumber \\
&\stackrel{H^{\otimes n}}{\mapsto}&  \frac{1}{2^{n}} \sum_{x,z=0} ^{x,z=2^{n}-1} (-1)^{f(x)+x.z}|z
\rangle, \label{eq:DJ}
\end{eqnarray}
where `$x.z$' refers to the scalar product between bitstrings ($x.z=x_{1}z_{1} \oplus
x_{2}z_{2}\oplus \ldots \oplus x_{n}z_{n}$). Consideration of the state $|z=0 \rangle$ reveals that
its amplitude is $1$ if the function is \emph{constant} and $0$ if the function is \emph{balanced},
so measuring along $|z=0 \rangle$ will allow perfect differentiation between the cases of the
function being balanced or constant. Achieving this feat classically would require examination of
half of the values of $f(x)$ plus one, for a total of $2^{n-1}+1$ queries and, therefore, an
exponential disadvantage as compared to the quantum algorithm.

If error is allowed, a classical algorithm needs far less queries to achieve a small error (see,
eg, \cite{jp:ln}). This algorithm progresses by looking at bits and as soon as a bit is observed
which is different to previous bits, the algorithm halts and returns the result `balanced', else it
terminates after $k$ bits examined (i.e., $k$ queries to the algorithm) and returns `constant'. The
only possibility of error, therefore, is when $k$ identical bits are observed (in which case the
result `constant' is returned) and the function is, in fact, balanced; the probability of this
occurring is just the chance of selecting $k$ identical objects from a sample of two species of
equal number.

\begin{equation}
P_{\mathrm{fail}} = 2p\frac{\left(\begin{array}{c}N/2
\\ k\end{array} \right)}{\left(\begin{array}{c}N \\ k\end{array} \right)} \label{eq:preskilfail}
\end{equation}
where $p$ is the probability that the function is actually balanced and $N=2^{n}$. If the constant
and balanced cases are equally likely and $k \ll N$,

\begin{equation}
P_{\mathrm{fail}} \simeq \frac{1}{2^{k}}, \label{eq:classpromise}
\end{equation}
and we can see that a comparatively small number of queries $k$ can produce an accurate answer.

\section{Breaking the Promise}\label{sec:brokprom}

We can imagine weakening the promise on the problem in the following way: introducing $y$ bit-flips
into the $N=2^{n}$ bit string that characterises the function $f$, at random positions, would
correspond to saying that the function is `nearly balanced' or `nearly constant', up to $y$
deviations from the promise. In this section, when we say `constant' we are actually referring to
`nearly constant', and similarly in the case of `balanced'.

\subsection{Deutsch-Jozsa algorithm with broken promise}\label{sec:qbrokprom}

The Deutsch-Jozsa (DJ) algorithm can be used unmodified on the new problem. If the function $f(x)$
is actually balanced, then in the coefficient of $|z=0\rangle$ in equation \ref{eq:DJ}, we no
longer see the cancellation caused by equal numbers of terms with $f(x)=1$ and $f(x)=0$ in the
index of the $-1$. We now have $(\frac{N}{2}+y)$ of $\pm 1$ in the phase of $|z=0 \rangle$ and
$(\frac{N}{2}-y)$ of $\mp 1$. There is, therefore, a non-zero coefficient of $|z=0\rangle$,
$\alpha_{0}$:

\begin{equation}
\alpha_{0} = \pm \frac{2y}{2^{n}},
\end{equation}
which leads to an error probability $P_{\mathrm{bal}}$ of (given that error in this case is
measuring $|z=0\rangle$):

\begin{equation}
P_{\mathrm{bal}} = y^{2}2^{2-2n}.   \label{eq:DJbalerror}
\end{equation}

 For $y$ weakenings in the promise on the \emph{actually constant} $n$-bit
function $f(x)$, we find that for the D-J algorithm we have:

\begin{equation}
\alpha_{0}= \pm \left( 1-\frac{2y}{2^n} \right).
\end{equation}

It is an error, in the case where $f(x)$ is constant, if we do \emph{do not} observe $|z=0\rangle$,
so our error probability in this case $P_{\mathrm{con}}$ is given by:

\begin{equation}
P_{\mathrm{con}}= y 2^{2-n} - y^{2}2^{2-2n}, \label{eq:DJconerror}
\end{equation}
and we see that the error is worse if the function is actually constant than if it is
balanced\footnote{This is because any redistribution of the outcome probabilities in the `constant'
case will give rise to error (because it gives rise to probability of observing $|z \neq 0\rangle$
, whereas the only redistribution of the outcome probabilities that will cause error in the
`balanced' case is that which specifically causes an increase in the probability of observing $|z=0
\rangle$}. Assuming that the probability that the function $f(x)$ is balanced is $p$, the error
probability $P_{\mathrm{fail}}$ from one query to the quantum algorithm is:

\begin{eqnarray}
P_{\mathrm{fail}} &=& pP_{\mathrm{bal}} + (1-p)P_{\mathrm{con}} \nonumber \\ &=& p y 2^{2-n} -p
y^{2}2^{2-2n} + (1-p)y^{2}2^{2-2n} \nonumber \\ &=& p y 2^{2-n}  + (1-2p)y^{2}2^{2-2n} ,
\end{eqnarray}
which, where each case is equally probable ($p=\frac{1}{2}$), is given by:

\begin{equation}
P_{\mathrm{fail}}= y 2^{1-n} \label{eq:quanterrpishalf}
\end{equation}
for a single query.

The chance of the quantum algorithm failing we assess  by majority decision over $k$ queries (for
direct comparison with van Dam's algorithm and the classical case); this is the chance that $\geq
\frac{k}{2}$ of the results return the wrong result (all evaluated on copies of the \emph{same}
function) and is given by:

\begin{equation}
P_{\mathrm{fail}} = \sum_{r=\frac{k}{2}}^{k} \left(\begin{array}{c}k \\ r\end{array}
\right)\left(pP_{\mathrm{bal}}^{r}(1-P_{\mathrm{bal}})^{k-r}+(1-p)P_{\mathrm{con}}^{r}(1-P_{\mathrm{con}})^{k-r}\right)
.    \label{eq:DJerror}
\end{equation}

We have that the mean failure probabilities for the balanced and constant strings are

\begin{eqnarray}
\mu_{\mathrm{bal}} &=& kP_{\mathrm{bal}} = \frac{4ky^{2}}{N^{2}} \\
\mu_{\mathrm{con}} &=& kP_{\mathrm{con}} = \frac{4ky}{N} \left(1-\frac{y}{N}\right)
\end{eqnarray}

and the variances are given by:

\begin{eqnarray}
\sigma^{2}_{\mathrm{bal}} &=& kP_{\mathrm{bal}}(1-P_{\mathrm{bal}}) = \frac{4ky^{2}}{N^{2}}\left(1-\frac{4y^{2}}{N^{2}}\right) \\
\sigma^{2}_{\mathrm{con}} &=& kP_{\mathrm{con}}(1-P_{\mathrm{con}}) = \frac{4ky}{N} \left(1 -
\frac{y}{N}\right)\left(1-\frac{4y}{N}+\frac{4y^{2}}{N^{2}}\right).
\end{eqnarray}

If we consider the limit where $y<<N$ and also that k is large enough to allow the binomial
distribution to be approximated by a normal distribution, we have that

\begin{equation} P_{\mathrm{fail}} \approx \frac{p}{2}\left[1-{\mathrm{erf}}\left(\frac{\frac{k}{2}-\mu_{\mathrm{bal}}}{\sigma_{\mathrm{bal}}\sqrt{2}}\right)\right]
+\frac{1-p}{2}\left[1-{\mathrm{erf}}\left(\frac{\frac{k}{2}
-\mu_{\mathrm{con}}}{\sigma_{\mathrm{con}}\sqrt{2}}\right)\right]
\end{equation}

Since the mean value for the constant probability is smaller, the term for the `near-constant'
string dominates the failure probability (since we are considering probabilities at the upper tail
end of the distribution). With $k$ large enough that we can use the approximation for the
complementary error function, ${\mathrm{erfc}}(t) \sim e^{-x^{2}}/x\sqrt{\pi}$, and with the
probabilities of actually having a balanced or constant function being equal, we have that

\beq P_{\mathrm{fail}} \approx \frac{1}{4}\left[1-{\mathrm{erf}}\left(\frac{\frac{k}{2}
-\mu_{\mathrm{con}}}{\sigma_{\mathrm{con}}\sqrt{2}}\right)\right]\eeq so that, taking the logarithm
of $P_{\mathrm{fail}}$ for ease of comparison with the other algorithms:
\begin{eqnarray}
\ln P_{\mathrm{fail}}&\approx &-\frac{k}{32}\left(\frac{y}{N}\right)^{-1}\frac{\left[1 -
\frac{8y}{N} \left(1-\frac{y}{N}\right)\right]^{2}}{\left(1 -
\frac{y}{N}\right)\left(1-\frac{4y}{N}+\frac{4y^{2}}{N^{2}}\right) }-\frac{1}{2}\ln k
+\frac{1}{2}\ln\frac{y}{N}\nonumber \\ &+&\frac{1}{2}\ln\left(1 -
\frac{y}{N}\right)\left(1-\frac{4y}{N}+\frac{4y^{2}}{N^{2}}\right)-\ln\left | 1- \frac{8y}{N}
\left(1-\frac{y}{N}\right)\right|+\frac{1}{2}\ln\frac{1}{\pi}. \ \ \ \ \ \ \label{eq:lnfailDJ}
\end{eqnarray}

\subsection{Classical algorithm with broken promise}\label{sec:cbrokprom}

Once we allow weakenings of the promise, the classical algorithm from section \ref{sec:DJpromise}
becomes particularly poor; whereas previously the algorithm only failed in the case that the
function was actually balanced, as the number of weakenings increases it becomes more probable that
a constant function will be described as balanced and it is this probability that dominates the
error probability relatively quickly. The existing algorithm is closely tied to the promise being
unbroken, but a new algorithm, based on sampling, is easily designed.

This algorithm queries $k$ values of the function and then makes a decision depending on the
relative numbers of bits in the sample, by counting the number of $0$s in the sample, $k_{0}$. The
proportion of $0$s in the sample is used as an estimator, $\hat{p}^{cl}_{0} = \frac{k_{0}}{k}$ for
the number of $0$s in the full $N$ bit string of $1$s and $0$s that describes $f$. The algorithm
then infers balanced if  $0.25 <\hat{p}^{cl}_{0}<0.75$, or constant otherwise. This estimator is
unbiased since its expected value $E(\hat{p}^{cl}_{0})=N_{0}/N$, the true proportion of zeros in
the string, independent of the degree of weakening. This algorithm avoids the severe failure in the
case of `actually constant' from which previous classical algorithm from section
\ref{sec:DJpromise} suffers.

This new classical algorithm fails if we have a constant (i.e., `weakened constant') string and yet
observe the number of $0$s, $k_{0}$, in the $k$ queries such that $\frac{k}{4} \leq k_{0} \leq
\frac{3k}{4}$, or if we have `balanced' and yet observe a number of $0$s, $k_{0}$, in the $k$
queries such that $k_{0} \leq \frac{k}{4}$ or $k_{0} \geq \frac{3k}{4}$ In both of these cases, the
wrong inference is made.

If we describe the event `observing $\frac{k}{4} \leq k_{0} \leq \frac{3k}{4}$' as $B$, the
property `string is balanced' as $A_{1}$ and the property `string is constant' as $A_{2}$, then we
can write an expression for the probability of algorithm failure, $P_{\mathrm{fail}}$, in terms of
the conditional probabilities of deciding on one case given that the other is actually the case:

\begin{equation}
P_{\mathrm{fail}}= p \left( 1 - \frac{1}{2}P(B|A_{1a}) - \frac{1}{2}P(B| A_{1b})\right) + (1-p)
\left( \frac{1}{2}P(B|A_{2a}) + \frac{1}{2}P(B| A_{2b}) \right), \label{eq:probclassfail}
\end{equation}
in which the subscript $a$ refers to the case when there is an excess of zeroes in the $N$ values
of $f(x)$ and $b$ refers to when there is an excess of ones and, as before, $p$ is the probability
that the function is actually balanced, and the classical probabilities can be well approximated by
the binomial distribution in the case of `large' query number, so that
\begin{eqnarray}
P(B|A_{1a}) &\approx& \sum ^{\frac{3k}{4}}_{k_{0}=\frac{k}{4}} \left(\begin{array}{c} k  \\ k_{0}
\end{array}\right)\left(\frac{1}{2}+\frac{y}{N}\right)^{k_{0}}\left(\frac{1}{2}-\frac{y}{N}\right)^{k-k_{0}}
\label{eq:case1ax} \\
P(B|A_{1b}) &\approx& \sum ^{\frac{3k}{4}}_{k_{0}=\frac{k}{4}} \left(\begin{array}{c} k  \\ k_{0}
\end{array}\right)\left(\frac{1}{2}-\frac{y}{N}\right)^{k_{0}}\left(\frac{1}{2}+\frac{y}{N}\right)^{k-k_{0}}
\label{eq:case1bx} \\
P(B|A_{2a}) &\approx& \sum ^{\frac{3k}{4}}_{k_{0}=\frac{k}{4}} \left(\begin{array}{c} k  \\ k_{0}
\end{array}\right)\left(1-\frac{y}{N}\right)^{k_{0}}\left(\frac{y}{N}\right)^{k-k_{0}}
\label{eq:case2ax} \\
P(B|A_{2b}) &\approx& \sum ^{\frac{3k}{4}}_{k_{0}=\frac{k}{4}} \left(\begin{array}{c} k  \\ k_{0}
\end{array}\right)\left(1-\frac{y}{N}\right)^{k-k_{0}}\left(\frac{y}{N}\right)^{k_{0}}
\label{eq:case2bx}  \\
\end{eqnarray}

 In the limit of large $k$ and $y/N < 1/4$, it is the `balanced' string that dominates the error probability.
 Asymptotically we can use the central limit theorem to approximate the cumulative probability as an error
 function,
 \begin{eqnarray}
  P(B|A_{1a}) &\approx& \frac{1}{2} {\mathrm{erf}}\left[\frac{\frac{3k}{4}-k\left(\frac{1}{2}-\frac{y}{N}\right)}
 {\sqrt{2k\left(\frac{1}{4}-\frac{y^{2}}{N^{2}}\right)}}\right]-
  \frac{1}{2} {\mathrm{erf}}\left[\frac{\frac{k}{4}-k\left(\frac{1}{2}-\frac{y}{N}\right)}
 {\sqrt{2k\left(\frac{1}{4}-\frac{y^{2}}{N^{2}}\right)}}\right]
 \\
 P(B|A_{1b}) &\approx& \frac{1}{2} {\mathrm{erf}}\left[\frac{\frac{3k}{4}-k\left(\frac{1}{2}+\frac{y}{N}\right)}
 {\sqrt{2k\left(\frac{1}{4}-\frac{y^{2}}{N^{2}}\right)}}\right]-
  \frac{1}{2} {\mathrm{erf}}\left[\frac{\frac{k}{4}-k\left(\frac{1}{2}+\frac{y}{N}\right)}
 {\sqrt{2k\left(\frac{1}{4}-\frac{y^{2}}{N^{2}}\right)}}\right].
 \end{eqnarray}

 $P_{\mathrm{fail}}$ is then given by:
 \begin{eqnarray}
 P_{\mathrm{fail}}&\approx& p(1- \frac{1}{2}(P(B|A_{1a}+P_{B|A_{1b}})))
 \end{eqnarray}
  Using the identity ${\mathrm{erf}}(-x) =-{\mathrm{erf}}(x)$ we find,
\beq
 P(B|A_{1a})+ P(B|A_{1b}) \approx {\mathrm{erf}}\left[\frac{\frac{k}{4}-\frac{ky}{N}}
 {\sqrt{2k\left(\frac{1}{4}-\frac{y^{2}}{N^{2}}\right)}}\right]+
 {\mathrm{erf}}\left[\frac{\frac{k}{4}+\frac{ky}{N}}{\sqrt{2k\left(\frac{1}{4}-\frac{y^{2}}{N^{2}}\right)}}\right]
 \eeq
 So that assuming $k$ is large and using the same approximation for ${\mathrm{erfc}}(x)$ as in the DJ case, we find
\begin{eqnarray}
\ln P_{\mathrm{fail}} &\approx&
-\frac{k}{8}\left[\frac{\left(1-\frac{4y}{N}\right)^{2}}{1-\frac{4y^{2}}{N^{2}}}\right]
-\frac{1}{2}\ln k -\frac{1}{2}\ln\left(1-\frac{4y}{N}\right)+\ln\left(1-\frac{4y^{2}}{N^{2}}\right)
+ \frac{1}{2}\ln\frac{8}{\pi}\ \ \ \ \ \ \ \label{eq:lnfailclassical}
\end{eqnarray}

\subsection{Comparison of Classical and Deutsch-Jozsa algorithms}

We can compare the classical and Deutsch-Jozsa algorithms, in the region where the approximations
hold, through examination of equations \ref{eq:lnfailDJ} and \ref{eq:lnfailclassical}. We are
primarily interested in the limit where $N \gg k \gg 0$. Equating the two expressions for $\ln
P_{\mathrm{fail}}$ and where $k$ is large, we have that equality in failure probability is
approximately achieved by solving:

\beq \frac{1}{32}\left(\frac{y}{N}\right)^{-1}\frac{\left[1 - \frac{8y}{N}
\left(1-\frac{y}{N}\right)\right]^{2}}{\left(1 -
\frac{y}{N}\right)\left(1-\frac{4y}{N}+\frac{4y^{2}}{N^{2}}\right) }
-\frac{1}{8}\left[\frac{\left(1-\frac{4y}{N}\right)^{2}}{1-\frac{4y^{2}}{N^{2}}}\right] =0.\eeq

We see that in this regime, the solution is independent of $k$ and that the degree of weakening at
which the DJ and classical algorithms yield the same failure probability tends to $y/N \approx
0.0973$. If the weakening of the promise is greater than this, the classical sampling approach
works better than the DJ-based approach.

\subsection{Quantum Oracle Interrogation}\label{sec:wvd}

Since reducing the strength of the promise is akin to making the problem less structured, we
compare the power of the DJ algorithm with that of an algorithm that is specifically tailored for
completely unstructured problems, the Quantum Oracle Interrogation algorithm of van Dam
\cite{{wvd:qo},{wvd:phd}} (WVD). In this case, there is no promise on the problem (i.e., all
possible $X$ are allowed).

As before, we use the fact that an $n$-bit function $f$ (i.e., $\{0,1\}^{n} \stackrel{f}{\mapsto}
\{0,1\}$ so that $N=2^{n}$) can be described by an $N$-bit string consisting of the values of
$f(x)$; this is represented as a state $|F\rangle=|f_{0}f_{1} \ldots f_{N-1} \rangle$, where
$f_{x}=f(x)$. An operator $A_{k}$ is introduced, the action of which depends on the Hamming weight
$||x||$ of the state $|x\rangle$ (ie the number of $1$s in the binary representation of $x$):

\begin{equation}
A_{k}|x \rangle |b \rangle = \left\{ \begin{array} {lc} |x \rangle | b \oplus (F.x)\rangle &\mathrm{if} \ ||x|| \leq k \\
| x \rangle|b \rangle &\mathrm{if} \ ||x|| > k \end{array} \right. .
\end{equation}

This operator requires at most $k$ queries, because the query complexity of $(F.x)$ is limited by
the Hamming weight $||x||$ of $|x \rangle$, and can be carried out on states in superposition. The
algorithm proceeds as follows:

\begin{itemize}
\item Prepare starting state $|\Psi_{k}\rangle \frac{1}{\sqrt{2}}\left(|0\rangle - |1 \rangle
\right)$
 ($| \Psi_{k} \rangle$ depends on $k$, as is explained below)
\item Act on state with $A_{k}$ \item Apply $H^{\otimes N}$ to the first $N$ quibits \item Discard
the ancilla qubit and measure the $N$ bits
\end{itemize}

For comparison with the previous two algorithms, we use the `approximate oracle interrogation'
version of WVD \cite{wvd:qo}, for which the starting state $|\Psi_{k}\rangle$ is prepared as
follows:

\begin{equation} |\Psi_{k}\rangle=\sum_{j=0}^{k}\frac{\alpha_{j}} {\sqrt{\left(\begin{array}{c}
  N \\ j \end{array}\right)}}\sum_{\substack{x\in\{0,1\}^{N} \\ ||x||=j}}|x\rangle. \end{equation}

Upon measurement, the observed state $|F'\rangle$ is `close' to $|F\rangle$, by which we mean that
we obtain an N-bit string of which $m$ bits are correct (that is, it shares $m$ bits with $F$, so
correctly represents the function $f$ for those $m$ values of $f(x)$) and the remainder incorrect,
where we don't know in which locations the correct bits are. The probability distribution for $m$
is given by
\begin{equation} P(m)=1-\frac{\left( \begin{array}{c}
  N \\
  m
\end{array}\right) }{ 2^{N}}\left|\sum_{j=0}^{k}\frac{\alpha_{j}}{ \sqrt{\left(\begin{array}{c}
  k \\ m \end{array}\right)}}{\mathcal K}_{j}(m;N)\right|^{2} \label{eq:pofmwvd}
\end{equation}
where ${\mathcal K}_{j}(m;N)$ is a Krawtchouk polynomial \cite{ms:ecc} given by
\begin{equation} {\mathcal K}_{j}(m;N)=\sum_{r=0}^{j}(-1)^{r}\left(\begin{array}{c}
  m \\ r \end{array}\right)\left(\begin{array}{c}
  N-m \\ j-r \end{array}\right).\end{equation}
We follow van Dam and use,
\begin{eqnarray} \alpha_{j}=& {1\over \sqrt[4]{k}}, \ \ \ & k-\sqrt{k}+1\le j\le k \nonumber \\
& 0, & {\rm otherwise.}\end{eqnarray} For large $N$, the optimal expected number of correct bits is
\begin{equation} E(m)_{opt}\approx{N\over 2}+\sqrt{k}\sqrt{N-k}.\label{expm}\end{equation} and
$P(m)$ becomes highly peaked around this value in the limit of large $N$ and $k$ as detailed in the
appendix.

Let us consider the function $f$ which is nearly balanced or constant with $N_{0}$ zeros. We tackle
the DJ problem in a similar way to the classical algorithm as discussed in section
\ref{sec:cbrokprom}, using an estimator for the proportion of zeros in the full string
($p_{0}=N_{0}/N$). In the WVD algorithm the analogous estimator is $\hat{p}_{0}=m_{0}/N$, where
$m_{0}$ is the number of zeros in the string after $k$ queries.  After $k$ queries one obtains m
correct bits of which $m_{0}^{*}$ are correct zeros. We assume that the correct bits are drawn at
random from the full population so that the distribution of correct zeros is, in the limit of large
$N$, given by

\beq P(m_{0}^{*}|m)\approx\left(\begin{array}{c} m \\ m_{0}^{*} \end{array}\right)\left(
\frac{N_{0}}{N} \right)^{m_{0} ^{*}}\left( 1-\frac{N_{0}}{N}\right) ^{m-m_{0} ^{*}} \eeq with
\begin{equation} E(m_{0}^{*}|m)={N_{0}\over N}m .\end{equation}

Considering each case separately in terms of the degree of weakening of the promise, $y/N$
(detailed in appendix) and assuming $k$ large
\begin{eqnarray}
\mu_{1a} &\approx& N\left[\frac{1}{2}+\frac{2y}{N}\sqrt{\frac{k}{N}\left(1-\frac{k}{N}\right)}\right]\label{eq:firstWVDexpectation}\\
\mu_{1b} &\approx& N\left[\frac{1}{2}-\frac{2y}{N}\sqrt{\frac{k}{N}\left(1-\frac{k}{N}\right)}\right]\\
\mu_{2a} &\approx& N\left[\frac{1}{2}+\left(1-\frac{2y}{N}\right)\sqrt{\frac{k}{N}\left(1-\frac{k}{N}\right)}\right]\\
\mu_{2b} &\approx&
N\left[\frac{1}{2}-\left(1-\frac{2y}{N}\right)\sqrt{\frac{k}{N}\left(1-\frac{k}{N}\right)}\right]\\
\sigma_{1}^{2}
&\approx&N\left(1-\frac{4y^{2}}{N^{2}}\right)\left[\frac{1}{2}+\sqrt{\frac{k}{N}\left(1-\frac{k}{N}\right)}\right]
\\
\sigma_{2}^{2}
&\approx&N\left(\frac{4y}{N}\right)\left(1-\frac{y}{N}\right)\left[\frac{1}{2}+\sqrt{\frac{k}{N}\left(1-\frac{k}{N}\right)}\right]\label{eq:secondWVDvariance}
\end{eqnarray}
We can see that $\mu$ is not an unbiased estimator for the number of zeros in the string for any of
these distributions, in contrast to the previous two examples, so an inference scheme based on $N/4
< m_{0} < 3N/4$ implying `balanced' will not be optimal where $y/N < 1/4$. We must therefore shift
the inference scheme to take this bias into account. The mid-points between the balanced and
constant expectation values are $N(\frac{1}{2}\pm
\frac{1}{2}\sqrt{\frac{k}{N}\left(1-\frac{k}{N}\right)})$. Let us therefore assume the following
scheme:
\begin{eqnarray}
\left|\frac{m_{0}}{N}-\frac{1}{2}\right|&<&\alpha\sqrt{\frac{k}{N}\left(1-\frac{k}{N}\right)}
\rightarrow
\mathrm{infer \ balanced}\\
\left|\frac{m_{0}}{N}-\frac{1}{2}\right|&>&\alpha\sqrt{\frac{k}{N}\left(1-\frac{k}{N}\right)}
\rightarrow \mathrm{infer \ constant}.  \end{eqnarray}

In the large $N$ limit, we can assume the central limit theorem holds for these binomial
distributions. The error probabilities then become:
\begin{eqnarray}
P_{\mathrm{con}} &\approx&
1-\frac{1}{2}{\mathrm{erf}}\left[\frac{\left(1-\alpha-\frac{2y}{N}\right)\sqrt{k\left(N-k\right)}}
{\sigma_{2}\sqrt{2}}\right]-\frac{1}{2}{\mathrm{erf}}\left[\frac{\left(1+\alpha-\frac{2y}{N}\right)\sqrt{k\left(N-k\right)}}
{\sigma_{2}\sqrt{2}}\right]\nonumber
\\
&\approx&\frac{1}{2}-\frac{1}{2}{\mathrm{erf}}\left[\frac{\left(1-\alpha-\frac{2y}{N}\right)\sqrt{k\left(N-k\right)}}
{\sigma_{2}\sqrt{2}}\right]
\\
P_{\mathrm{bal}}&\approx&
1-\frac{1}{2}{\mathrm{erf}}\left[\frac{\left(\alpha+\frac{2y}{N}\right)\sqrt{k\left(N-k\right)}}
{\sigma_{1}\sqrt{2}}\right]-\frac{1}{2}{\mathrm{erf}}\left[\frac{\left(\alpha-\frac{2y}{N}\right)\sqrt{k\left(N-k\right)}}
{\sigma_{1}\sqrt{2}}\right] \nonumber
\\
&\approx&\frac{1}{2}-\frac{1}{2}{\mathrm{erf}}\left[\frac{\left(\alpha-\frac{2y}{N}\right)\sqrt{k\left(N-k\right)}}
{\sigma_{1}\sqrt{2}}\right].
\end{eqnarray}

We now want to calculate the value of $\alpha$ to minimise $P_{\mathrm{fail}}$, so that using the
fact that $\frac{d }{dx}{\mathrm{erf}}(x)=e^{-x^{2}}$, and with $k,N \gg 0$, we have that this
value is given by \beq \alpha \approx
1+\frac{2y}{N}-\frac{4y^{2}}{N^{2}}-2\sqrt{\frac{y}{N}\left(1-\frac{y}{N}\right)\left(1-\frac{4y^{2}}{N^{2}}\right)}
\label{eq:optimal alpha}\eeq so that knowledge of the quantity $y/N$ is required to optimise this
inference rule, and it is this optimal value that we shall be indicating by $\alpha$ in what
follows.

Finally, we have that

\bea
P_{\mathrm{fail}}&\approx&\frac{1}{2}-\frac{1}{4}{\mathrm{erf}}\left[\sqrt{k}\frac{\left(\alpha-\frac{2y}{N}\right)}{\sqrt{\left(1-\frac{4y^{2}}{N^{2}}\right)}}f(N,k)\right]-
\frac{1}{4}{\mathrm{erf}}\left[\sqrt{k}\frac{\left(\alpha-\frac{2y}{N}\right)}{\sqrt{\left(\frac{4y}{N}-\frac{4y^{2}}{N^{2}}\right)}}f(N,k)\right],\nonumber\\
\eea where

\beq f(N,k)^{2} = \frac{(1-\frac{k}{N})}{\left(1+2\sqrt{\frac{k}{N}(1-\frac{k}{N})}\right)}. \eeq
In the limit of $k>>0$ and $y/N<<1$ we find
\begin{eqnarray}
P_{\mathrm{fail}}&\approx&\frac{1}{4}\left[1-{\mathrm{erf}}\left(\sqrt{k}\frac{\left(\alpha-\frac{2y}{N}\right)}{\sqrt{\left(1-\frac{4y^{2}}{N^{2}}\right)}}f(N,k)\right)\right]
\\
\ln
P_{\mathrm{fail}}&\approx&-k\frac{\left(\alpha-\frac{2y}{N}\right)^{2}}{\left(1-\frac{4y^{2}}{N^{2}}\right)}f(N,k)^{2}-\mathcal{O}\left(\ln
k\right). \label{eq:lnfailWVD}\end{eqnarray}

\subsection{Comparison of DJ and WVD algorithms}

Comparing equations \ref{eq:lnfailDJ} and \ref{eq:lnfailWVD}, in the limit of $N \gg k \gg 0$, we
find that the error probabilities from the DJ and WVD algorithms are approximately equal when

\beq \frac{k}{32}\left(\frac{y}{N}\right)^{-1}\frac{\left[1 - \frac{8y}{N}
\left(1-\frac{y}{N}\right)\right]^{2}}{\left(1 -
\frac{y}{N}\right)\left(1-\frac{4y}{N}+\frac{4y^{2}}{N^{2}}\right) }
-k\frac{\left(\alpha-\frac{2y}{N}\right)^{2}}{\left(1-\frac{4y^{2}}{N^{2}}\right)} = 0 \eeq with
$\alpha$ given by equation \ref{eq:optimal alpha} so that the DJ and WVD algorithms in this limit
share the same error probability when $y/N \approx 0.0499$. For values of $y/N$ greater than this,
the WVD algorithm is superior in correctly deciding whether or not the function in question is
balanced or constant, given the assumption of large $N$ and $k$. Furthermore, we note that in the
region in which the given approximations hold, the classical algorithm is never better than the WVD
algorithm.

\section{Conclusion}

\begin{figure}[htbp]
\centerline{\includegraphics[width=8cm]{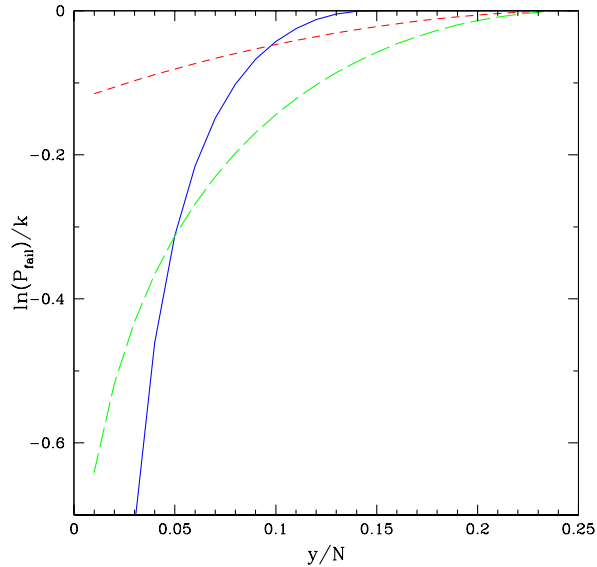}} \caption{\label{fig:wvdfig5} $\ln
P_{\mathrm{fail}}/k$ against degree of weakening $y/N$, for DJ algorithm (unbroken line), WVD
algorithm (long-dashed line) and classical algorithm (short-dashed line)}
\end{figure}

Figures \ref{fig:wvdfig5}, in which the relative performances of the three algorithms considered
here are compared, illustrates that the problem is still structured; the DJ algorithm is successful
in solving the problem but loses power as the promise becomes more broken. The WVD algorithm, which
is designed for determining functions that have no promise on their nature, can be adapted to
perform well even for relatively small weakenings of the promise, and becomes superior to the DJ
algorithm when more than about a twentieth of the bits defining the function under examination are
flipped.

We have found that the DJ algorithm is surprisingly robust when we consider breaks in the promise,
with the failure probabilities driven predominantly the chance of incorrectly inferring a constant
function. The speed-up in comparison to a classical sampling algorithm, in terms of queries to
achieve the same result, is obviously reduced over the case of the unbroken promise, but there is
still an advantage for smaller weakenings of the promise.

When compared against the quantum algorithm tailored to unstructured problems, WVD, the DJ
algorithm outperforms it for low degree of weakening of the promise. The adaptation of the
inference protocol for the WVD algorithm, based on the distribution of zeroes in the function
string, is crucial in minimising the probability of incorrect decision, particularly for the
smaller weakenings of the promise and thus the degree of the weakening of the promise needs to be
known to optimise the success probability. The DJ algorithm, designed for the unbroken promise,
nevertheless retains much of its advantage in the case where the promise is not greatly weakened
and does not require modification according to the degree of weakening of the promise.

Given access to all three algorithms, the DJ algorithm would be preferred if the weakening of the
promise was described by $y/N < .0499$ and the WVD would be preferred if the weakening was greater
than this, given the assumption of large $N$, $k$ and $N \gg k$. The classical algorithm would
never be preferred in this regime.

\section*{Acknowledgements}

We thank Peter L Knight for useful discussions and Wim van Dam for communications including sharing
details of his calculations on the success probability of his algorithm from his PhD thesis. AB
acknowledges financial support from the Engineering and Physical Sciences Research Council (EPSRC).

\section*{Appendix} \label{sec:appA}

We approach the problem by first calculating the expectation and variance of $m$ as a function of
$k$ and $N$ in the large $N$ (and $k$) limit. We follow van Dam and consider a perfectly constant
string of N bits all of which are zero. We then obtain the expectation and variance for the number
of 1's in the string, $t$, produced by $k$ queries and this result is then good, in fact, for the
expected number of incorrect bits for any string under examination with the WVD algorithm.

Van Dam derived $E(t|N,k)$; we reproduce this result initially for completeness and then evaluate
$E(t^{2}|N,k)$. The probability distribution for the number of $1$s, $t$, in the output string from
the WVD algorithm is given by
\begin{eqnarray}
P(t|N,k) &=&\sum_{i,j=0}^{k}\alpha_{i}\alpha_{j}^{*}\gamma_{ij}(N,t)\\
\gamma_{ij}(N,t) &=& \frac{1}{2^{N}}\left(\begin{array}{c}N\\
t\end{array}\right)\frac{{\mathcal K}_{i}(t;N)
{\mathcal K}_{j}(t;N)}{\sqrt{\left(\begin{array}{c}N\\
i\end{array}\right)\left(\begin{array}{c}N\\
j\end{array}\right)}}
\end{eqnarray}

We are interested in calculating the first and second moments of $t$ i.e. $E(t|N,k)$ and
$E(t^{2}|N,k)$ for which we introduce the following notation:
\begin{eqnarray}
\beta_{ij}^{(n)}(N,t)&=& \sum_{t=0}^{N}t^{n}\gamma_{ij}(N,t) \\
E(t^{n}|N,k) &=&\sum_{i,j=0}^{k}\alpha_{i}\alpha_{j}^{*}\beta_{ij}^{(n)}
\end{eqnarray}

We use the following Krawtchouk polynomial identities: the orthogonality relation of the Krawtchouk
polynomials \cite{ms:ecc}
\begin{eqnarray}
\sum_{t=0}^{N}\left(
\begin{array}{c}  N \\ t\end{array}\right){\mathcal K}_{i}(t;N){\mathcal K}_{j}(t;N) = 2^{N}\left(\begin{array}{c}  N \\
j\end{array}\right)\delta_{ij}\label{Krawortho}
\end{eqnarray}
and the three-term recursion relation \cite{ms:ecc}
\begin{equation}
(N-2t){\mathcal K}_{j}(t;N) \equiv (j+1) {\mathcal K}_{j+1}(t;N) + (N-j+1){\mathcal K}_{j-1}(t;N)
\label{Krawrec}
\end{equation}
Multiplying \ref{Krawrec} by $2^{-N} \left(\begin{array}{c}N\\
t\end{array}\right)\mathcal{K}_{i}(t;N)/\sqrt{\left(\begin{array}{c}N\\
i\end{array}\right)\left(\begin{array}{c}N\\
j\end{array}\right)}$, summing over $t$ and invoking the orthogonality relation \ref{Krawortho} we
find

\beq \beta_{ij}^{(1)} =
\frac{N}{2}\delta_{i,j}-\frac{1}{2}\sqrt{i(N-i+1)}\delta_{i,j+1}-\frac{1}{2}\sqrt{(i+1)(N-i)}\delta_{i,j-1}
\eeq so that taking WVD's weightings $\alpha_{i}=k^{\frac{1}{4}}$ for $k-\sqrt{k}\le j\le k$ and 0
otherwise,
\begin{eqnarray}
E(t|N,k)
&=&\sum_{i=0}\frac{N}{2}|\alpha_{i}|^{2}-\frac{1}{2}\sqrt{(i+1)(N-i)}(\alpha_{i+1}^{*}\alpha_{i}+\alpha_{i}^{*}\alpha_{i+1}) \\
&=&\frac{N}{2}-\frac{1}{\sqrt{k}}\sum_{j=k-\sqrt{k}}^{k}\sqrt{(i+1)(N-i)}
\end{eqnarray}
In the limit that $k,N>>1$ then each of the $\sqrt{k}$ terms in the sum is approximately of the
order $\sim\sqrt{k(N-k)}$ and we can approximately write
\begin{eqnarray}
E(t|N,k)&\approx&\frac{N}{2}-\sqrt{k(N-k)} +{\mathcal O}(\sqrt{N})
\end{eqnarray}
 In order to calculate the variance we need to evaluate $E(t^{2}|N,k)$.
We evaluate $\beta^{(2)}_{ij}$ by squaring the identity \ref{Krawrec} and then, again, multiplying
by $2^{-N}\left(\begin{array}{c}N\\t\end{array}\right)\mathcal{K}_{i}(t;N)$, summing over $t$ and
invoking the orthogonality relation \ref{Krawortho}
\begin{eqnarray}
LHS &\equiv &
N^{2}\left(\begin{array}{c}N\\i\end{array}\right)\delta_{i,j}-4N\left(\begin{array}{c}N\\i\end{array}\right)\left[
\frac{N}{2}\delta_{i,j}-\frac{1}{2}\sqrt{i(N-i+1)}\delta_{i,j+1} \nonumber\right.\\
&&\left.-\frac{1}{2}\sqrt{(i+1)(N-i)}\delta_{i,j-1}\right]
+\frac{4}{2^{N}}\sum_{t=0}^{N}t^{2}\left(\begin{array}{c}N\\t\end{array}\right)K_{i}(t,N)K_{j}(t,N) \\
RHS &\equiv& (i+1)(j+1) \left(\begin{array}{c}  N \\
i+1\end{array}\right)\delta_{ij} + (i+1)(N-j+1)
\left(\begin{array}{c}  N \\
i+1\end{array}\right)\delta_{i+1,j-1}\nonumber \\ &&+(N-i+1)(j+1)\left(\begin{array}{c}  N \\
i-1\end{array}\right)\delta_{i-1,j+1}+(N-i+1)(N-j+1)\left(\begin{array}{c}  N \\
i-1\end{array}\right)\delta_{ij}  \nonumber\\
\end{eqnarray}

It is only the diagonal terms and the off-diagonal terms differing by $|i-j|=1$ and $|i-j|=2$ that
are non-zero and we find that:

\begin{equation}\begin{array}{lcll}
\beta^{(2)}_{ij} &=& \frac{1}{4}((i+1)(N-i)+i(N-i+1)+N^{2}) &\mathrm{if \ } i=j\\
&=&-\frac{1}{2}N\sqrt{i(N-i+1)} &\mathrm{if \ } i=j+1\\
&=&-\frac{1}{2}N\sqrt{(i+1)(N-i)}& \mathrm{if \ } i=j-1\\
&=& \frac{1}{4}\sqrt{(N-i)(N-i-1)(i+1)(i+2)}\ &\mathrm{if \ } i+1=j-1\\
&=& \frac{1}{4}\sqrt{(N-i+2)(N-i+1)(i-1)i}\ &\mathrm{if \ } i-1=j+1 \\
&=& 0 &\mathrm{otherwise} \end{array}
\end{equation}so, therefore, assuming WVD's choice of weighting $\alpha_{i}$
\begin{eqnarray}
E\left(t^{2}|N,k\right)
&=&\frac{1}{4\sqrt{k}}\sum_{i=k-\sqrt{k}}^{k}\left(i+1\right)\left(N-i\right)
+i\left(N-i-1)\right)+N^{2} \nonumber \\ &&
-\frac{1}{2}\frac{2}{\sqrt{k}}\sum_{i=k-\sqrt{k}}^{k-2}\sqrt{\left(i+1\right)\left(N-i)\right)} \nonumber\\
\nonumber \\ &&
+\frac{1}{4}\frac{2}{\sqrt{k}}\sum_{i=k-\sqrt{k}}^{k-2}\sqrt{\left(N-i\right)\left(N-i-1)\right)
\left(i+1\right)\left(i+2\right)} \nonumber\\
&\stackrel{k,N>>1}{\approx}&\frac{N^{2}}{4}+k\left(N-k\right)-N\sqrt{k\left(N-k\right)} +
\mathcal{O}(N)
\end{eqnarray} and recalling that $m=N-t$, we find in the limit of $k,N >>1$
\begin{eqnarray}
E(m) &\approx& N-E(t) =\frac{N}{2}+\sqrt{k(N-k)}\label{eq:expectationformWVD}
\\
 Var\left(m\right) &=& Var\left(t\right) = E\left(t^{2}\right)-
\left[E\left(t\right)\right]^{2}\nonumber
\\&\approx&\frac{N^{2}}{4}+k\left(N-k\right)-N\sqrt{k\left(N-k\right)}-\left[\frac{N}{2}-\sqrt{k(N-k)}
\right]^{2}\nonumber \\
&\approx& 0 \label{eq:varianceis0WVD}.
\end{eqnarray}
We can therefore assume that on exiting the quantum query algorithm we know that effectively a
fixed number of bits, $m=\frac{N}{2}+\sqrt{k(N-k)}$ are correct. We assume that the algorithm knows
nothing of the nature of the bits and therefore correctly or incorrectly ascertains a given bit's
value randomly so that the number of correct zeros is given by, in limit of large $N$, \beq
P(m_{0}^{*}|m)=\left(\begin{array}{c} m \\ m_{0}^{*} \end{array}\right)\left( \frac{N_{0}}{N}
\right)^{m_{0} ^{*}}\left( \frac{N-N_{0}}{N}\right) ^{m-m_{0} ^{*}} \eeq with
\begin{eqnarray}
E(m_{0}^{*}) &=& \frac{N_{0}}{N} m
=\frac{N_{0}}{N}\left(\frac{N}{2}+\sqrt{k(N-k)}\right)\\
Var(m_{0}^{*})
&=&\frac{N_{0}}{N}\left(1-\frac{N_{0}}{N}\right)\left(\frac{N}{2}+\sqrt{k(N-k)}\right)
\end{eqnarray}
so that, given $m_{0} = N-N_{0} - m +2m_{0} ^{*}$ and using equations \ref{eq:expectationformWVD}
and \ref{eq:varianceis0WVD}
\begin{eqnarray}
\mu = E(m_{0}) &\approx& N\left[\frac{1}{2}+\left(2\frac{N_{0}}{N}-1\right)\sqrt{\frac{k}{N}\left(1-\frac{k}{N}\right)}\right]\\
\sigma^{2} =Var(m_{0})
&\approx&4N_{0}\left(1-\frac{N_{0}}{N}\right)\left[\frac{1}{2}+\sqrt{\frac{k}{N}\left(1-\frac{k}{N}\right)}\right]
\end{eqnarray}

Considering each case separately in terms of the degree of weakening of the promise and
substituting for $N_{0}$ in terms of $y, N$, we obtain equations
{\ref{eq:firstWVDexpectation}-\ref{eq:secondWVDvariance}}.

\end{document}